\documentclass{ws-procs9x6}
\usepackage{epsf2}
\usepackage{graphicx}
\newcommand{\mps}{m_{PS}}
\newcommand{\be}{\begin{equation}}
\newcommand{\ee}{\end{equation}}

\begin{document}

\title{
\vspace{-7.8em}
\parbox{80mm}{\hbox to \hsize
{\hss \vspace{-1.0em}\normalsize LTH 602}}
\parbox{80mm}{\vspace{-3.3em}
\hbox to \hsize
{\hss \normalsize Edinburgh 2003/21}}
\parbox{130mm}{\vspace{-1.2em}\hbox to \hsize
{\hss \normalsize LU-ITP 2003/020}}
\parbox{130mm}{\vspace{-7.8em}\hbox to \hsize
{\hss \normalsize HU-EP-03/67}}
\parbox{130mm}{\vspace{-8.1em}\hbox to \hsize
{\hss \normalsize DESY 03-190}}\\
Calculation of Finite Size Effects on the Nucleon Mass in unquenched QCD using
Chiral Perturbation Theory\footnote{presented by \uppercase{A}. \uppercase{A}li \uppercase{K}han}}

\author{A.~Ali Khan$^a$, T.~Bakeyev$^b$, M.~G\"ockeler$^{c,d}$, T.R.~Hemmert$^e$, R.~Horsley$^f$, 
A.C.~Irving$^g$, D.~Pleiter$^h$, P.E.L.~Rakow$^g$, G.~Schierholz$^{h,i}$ and H.~St\"uben$^j$}

\address{$^A$Institut f\"ur Physik, Humboldt-Universit\"at 
zu Berlin,  12489 Berlin, Germany, \\
$^B$Joint Institute for Nuclear Research, 141980 Dubna, Russia, \\
$^C$Institut f\"ur Theoretische Physik, Universit\"at 
Leipzig, 04109 Leipzig, Germany, \\
$^D$Institut f\"ur Theoretische 
Physik, Universit\"at Regensburg, 93040 Regensburg, Germany, \\
$^E$Physik-Department, Theoretische Physik T39, TU M\"unchen,
85747 Garching, Germany, \\
$^F$School of Physics, The University of Edinburgh, Edinburgh EH9 3JZ, UK, \\
$^G$Theoretical Physics Division, Department of 
Mathematical Sciences, University of Liverpool,  
Liverpool L69 3BX, UK, \\
$^H$John von Neumann-Institut f\"ur Computing NIC, 
15738 Zeuthen, Germany, \\
$^I$Deutsches Elektronen-Synchrotron  DESY, 22603 Hamburg, Germany, 
$^J$Konrad-Zuse-Zentrum f\"ur Informationstechnik
Berlin, 14195 Berlin, Germany \\
(QCDSF and UKQCD Collaborations)
}       





\maketitle

\abstracts{
The finite  size effects on nucleon masses are calculated in relativistic
chiral perturbation theory. Results are compared with two-flavor lattice results.}

\section{Introduction}
Finite size effects, in particular on dynamical configurations, can be a serious impediment
to precision lattice calculations of hadron masses and matrix elements. 
On the lattice sizes used in production runs, the nucleons and pions are the
relevant degrees of freedom  for understanding the finite size effects on the nucleon mass. 
We calculate these effects here using  two-flavor relativistic baryon chiral perturbation theory ($\chi PT$) 
at $O(q^3)$ (one loop).

\begin{figure}[t]
\epsfysize=6.5cm \epsfbox{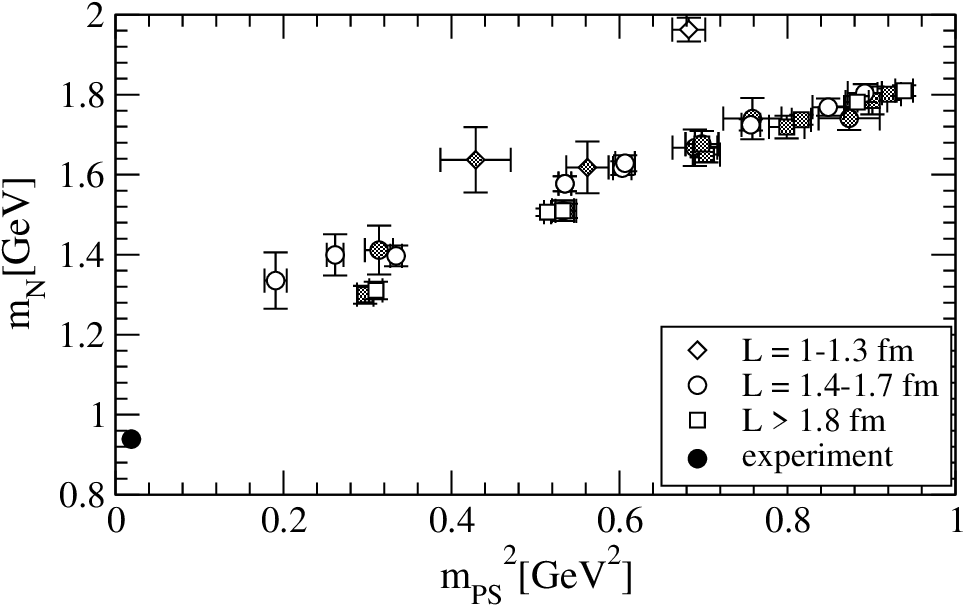}
\caption{Nucleon masses from QCDSF and UKQCD (white symbols) and JLQCD (gray symbols).
}
\label{fig:oaimproved}
\end{figure}

This study of nucleon masses is based on QCDSF and UKQCD unquenched ($N_f = 2$) lattice data
using a Wilson gauge action and two flavors of non-perturbatively 
$O(a)$-improved Wilson fermions ($a$ denotes the lattice spacing).
The corresponding pion masses are in the interval $0.45-1$ GeV.
Valence and sea quark masses are equal. Lattice sizes are $1-2.2$ fm.
The scale is set with $r_0$, using the physical value
$r_0 = 0.5$ fm $\simeq 1/(395 \mathrm{MeV}$). 
JLQCD has undertaken a study with the same lattice actions and range of simulation 
parameters on varying lattice sizes~\cite{JLQCD02}. In the comparison shown in 
Fig.~\ref{fig:oaimproved} one can observe that masses from lattices smaller than
$1.8$ fm are consistently larger than the results from larger lattices. 

The $a \rightarrow 0$ limit was not performed. Since discretization effects in 
unquenched simulations could be quite large~\cite{sommer} we have to keep an eye on the 
discretization errors in our results. Studying the variation of nucleon masses with $a$ at a fixed
lattice size we find that between two data points on approximately the same 
volume and at the same pion mass, but with $a$ varying by $\sim 30\%$, the nucleon mass remains
unchanged within the statistical errors. 
Moreover, we compare our results from large lattices ($L \geq 1.8$ fm) 
with two CP-PACS data sets~\cite{cppacs01} with renormalization group improved gauge fields and 
tree-level tadpole-improved clover quarks, $a^{-1}\approx 1.5$ and 2 GeV respectively 
and $L \geq 2.5$ fm, and find that they agree.
\section{$\chi PT$ formalism}
The one-loop contribution at $O(q^3)$ is generated by the $O(q^1)$ Lagrangian ${\mathcal L}_N^{(1)}$:
\be
{\mathcal L}^{(1)}_N =  \bar{\Psi}\left(i\gamma_\mu D^\mu - m_0\right)\Psi + 
\frac{1}{2} g_A\bar{\Psi} \gamma_\mu \gamma_5 u^\mu \Psi ,
\ee
with $D_\mu = \partial_\mu + \frac{1}{2}[u^\dagger,\partial_\mu u]$, 
$u_\mu = iu^\dagger \partial_\mu U u^\dagger$ and $u^2 = U$.
We use the infrared regularization 
 scheme~\cite{becher99}.
To compute the renormalized nucleon mass $m_N$, we include the tree-level $O(q^2)$ 
term $-4c_1\mps^2$ and an additional contribution of the form $e_1\mps^4$, which is
needed for renormalization of the relativistic $O(q^3)$ result, although formally
only entering at $O(q^4)$.
The one-loop correction to $m_0$ is~\cite{proc_future}:
\begin{eqnarray} 
 m_N &=& m_0 - 4c_1 \mps^2 + \left[e_1^r(\lambda) + \frac{3g_A^2}{64\pi^2F_\pi^2m_0}
\left(1-2\ln\frac{m_{PS}}{\lambda}\right)\right]m_{PS}^4 \nonumber \\
&- & \frac{3g_A^2}{16\pi^2F_\pi^2}m_{PS}^3\sqrt{1-\frac{m_{PS}^2}{4m_0^2}} 
\left[\frac{\pi}{2} + \arctan \frac{m_{PS}^2}{\sqrt{4\mps^2 m_0^2 - \mps^4}}\right]. 
 \label{eq:arctan}
\end{eqnarray}
$e_1^r(\lambda)$ denotes the renormalized $e_1$. 
The renormalization procedure is detailed in~\cite{proc_future}. 
We employ $\lambda = 1 \mathrm{GeV}$, $g_A =  1.2$~\cite{proc03}, 
and $F_\pi = 92.4$ MeV.
\begin{figure}[t]
\epsfysize=6.0cm \epsfbox{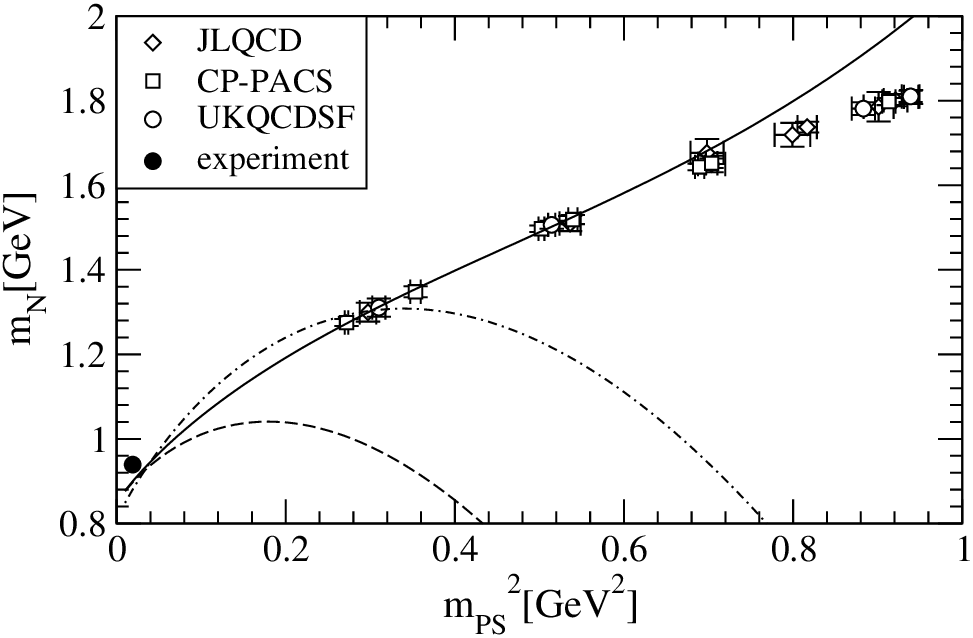}
\vspace{-5pt}
\caption{Nucleon masses on large lattices. The solid curve represents a fit with 
relativistic $\chi PT$, the dashed curve the NR limit using the same values of
$m_0$ and $c_1$, and the dot-dashed curve the NR limit using $m_0 = 0.81$ GeV
and $c_1 = -1.1$ GeV$^{-1}$.
}
\label{fig:largelat}
\vspace{-10pt}
\end{figure}

The nucleon mass in the chiral limit, $m_0$, and  the values of $c_1$ and $e_1^r$ are determined 
by a fit to six lattice data points at the smallest
masses. The fit is shown in Fig.~\ref{fig:largelat}. Relativistic $\chi PT$
describes the data correctly for pion masses up to $\sim 700$ MeV.
For the values of the parameters, we find
$m_0 = 0.85(14)\mathrm{GeV}$, $c_1=-0.80(18)\mathrm{GeV}^{-1}$ and $e_1^r(1
\,\mathrm{GeV})=2.8(1.1) \mathrm{GeV}^{-3}$.  
These results differ slightly from the ones quoted in~\cite{proc_future}
due to their fitting to a larger set of lattice points and using the value $g_A=1.267$.

In the limit $\mps \ll m_N$, keeping only terms up to $O(\mps^3)$ of 
Eq.~(\ref{eq:arctan}), one obtains non-relativistic (NR) $\chi PT$. In Fig.~\ref{fig:largelat},
we also plot NR results. We find that they approximate the data and the relativistic curve only for rather
small pion masses. A method to extend the validity of the NR 
approximation  to higher momentum scales within a cutoff scheme is described in~\cite{bernard03}.
The breakdown of the non-relativistic theory around $\mps =400$ MeV as shown in the plots
can be avoided if momentum modes above a cutoff $\Lambda_\chi$  are absorbed in local counterterms
(e.g. see the discussion in~\cite{bernard03}). Then the
nonrelativistic $O(q^3)$ calculation actually agrees quite well with the corresponding relativistic 
result up to $\mps=600$ MeV.
\section{Finite Size Effects}
We determine the finite size effect from the one-loop $O(q^3)$ contribution 
to the self-energy. Putting the external nucleon line on-shell,
the self-energy is given by~\cite{becher99}
\begin{eqnarray}
\! \Sigma(q\!\!\slash=m_0)\! & = & \!
\frac{3g_A^2m_0\mps^2}{2F_\pi^2}\int_0^\infty \!\!\! dx \!
\int\!\!\frac{d^4p}{(2\pi)^4}
\left[p^2+m_0^2x^2 + \mps^2(1-x)\right]^{-2}  
\end{eqnarray}
in Euclidean space. 
The temporal extent of the lattice is taken  to be infinite.
We define
$ \delta  = \frac{1}{m_0}
\left(\Sigma(q\!\!\!\slash=m_0,L) - \Sigma(q\!\!\!\slash=m_0,\infty)\right)$,
and using~\cite{hasenfratz90} express it  as an integral over Bessel functions:
\begin{eqnarray}
 \delta & = & 
\frac{3g_A^2\mps^2}{16\pi^2F_\pi^2}
  \int_0^\infty dx   \sum_{\vec{n}\neq 0} 
 K_{0}\left(L|\vec{n}|\sqrt{m_0^2x^2+\mps^2(1-x)}\right).
\end{eqnarray}
Then we calculate the finite size behavior of the nucleon mass:
\be
 m_N(L) =  (1+\delta)m_N(\infty). \label{eq:delta}
\ee
We first extrapolate to $m_N(\infty)$ by using the lattice result from the largest 
available lattice as input to~Eq.(\ref{eq:delta}), and then determine
$m_N(L)$ at finite lattice extent using the formula~Eq.(\ref{eq:delta}).
\begin{figure}[t]
\epsfysize=6.0cm \epsfbox{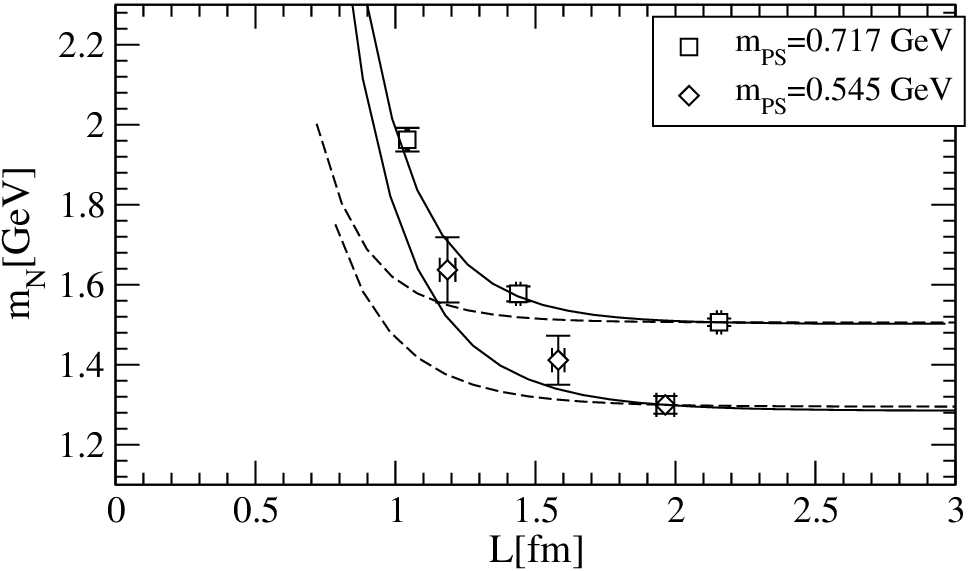}
\vspace{-5pt}
\caption{$m_N$ as a function of size on the lattice and in
$\chi PT$. Data sets correspond to a fixed $(\beta,\kappa)$ value and 
are labeled by the pion mass on the largest volume. The dashed lines show
the results from NR $\chi PT$.}
\label{fig:chPTfinitevol}
\vspace{-10pt}
\end{figure}
In Fig.~\ref{fig:chPTfinitevol} we compare lattice nucleon masses from 
various volumes with both relativistic and NR $\chi PT$  at $O(q^3)$.
Relativistic $\chi PT$ describes 
the lattice data very  well up to pion masses $\sim 700$ MeV. We find that for the finite size effect 
to be $< 1\%$,
$L$ should be $\geq 1.9$ fm at the smallest pion mass shown in the plot.

The finite size effect calculated in the non-relativistic formalism (for discussion see 
also~\cite{alikhan02}) is considerably
smaller compared to the relativistic result and the lattice data. On lattices of $1-2$ fm size
it is $\sim 30-40\%$ of the relativistic effect.

\noindent {\bf Acknowledgements} \\
This work is supported in part by the Deutsche Forschungsgemeinschaft. 
The simulations were done on the APEmille at NIC (Zeuthen),
  the Hitachi SR8000 at LRZ (Munich) and the Cray T3E at EPCC (Edinburgh)
  and NIC (J\"{u}lich). We thank all institutions for their support.

\end{document}